\input{aipcheck}

\documentclass[
    ,final            
  ]
  {aipproc}

\layoutstyle{8x11single}

\begin{document}

\title{Symmetry-preserving contact interaction model 
for heavy-light mesons}

\classification{11.10.St, 11.15.Tk, 11.30.-j, 14.40.Df, 14.40.Lb, 24.85.+p}
\keywords{Regularization of ultraviolet divergences, Ward-Green-Takahashi identities,
Dyson-Schwinger and Bethe-Salpeter equations}

\author{F.~E.~Serna}{
  address={Instituto de F\'{\i}sica Te\'{o}rica, Universidade Estadual
Paulista \\
Rua Dr. Bento Teobaldo Ferraz, 271 - Bloco II, 01140-070 S\~{a}o Paulo, SP, Brazil}
}

\author{M.~A.~Brito}{
}

\author{G.~Krein}{
}

\begin{abstract}
We use a symmetry-preserving regularization method of ultraviolet divergences in a 
vector-vector contact interaction model for low-energy QCD. The contact interaction 
is a representation of nonperturbative kernels used Dyson-Schwinger and Bethe-Salpeter 
equations. The regularization method is based on a 
subtraction scheme that avoids standard steps in the evaluation of divergent integrals 
that invariably lead to symmetry violation.  Aiming at the study of heavy-light mesons, we
have implemented the method to the pseudoscalar $\pi$ and $K$ mesons. We have solved the 
Dyson-Schwinger equation for the $u,d$ and $s$ quark propagators, and obtained 
the bound-state Bethe-Salpeter amplitudes in a way that the Ward-Green-Takahashi
identities reflecting global symmetries of the model are satisfied for arbitrary routing 
of the momenta running in loop integrals.  
\end{abstract}

\maketitle

\section{Introduction}
Contact fermionic interactions find widespread applications in hadron physics with 
Nambu--Jona-Lasinio (NJL)~\cite{Nambu:1961tp} type of models. A great deal of qualitative 
insight on the phenomenon of hadron mass generation via dynamical chiral symmetry breaking
(D$\chi$SB) and the role of the pions as the associated (quasi) Goldstone bosons has been 
gleaned from such models  -- for reviews, see Refs.~\cite{{Vogl:1991qt},{Klevansky:1992qe},
{Hatsuda:1994pi},{Bijnens:1995ww}}. The lack of confinement and nonrenormalizabilty 
are the major weaknesses of contact fermionic models. The nonrenormalizability carries with 
it the danger of introducing gross violations of global symmetries due to the regularization 
procedure; ambiguities arising from momentum shifts in divergent integrals are the main cause 
of the problems. Despite well known, practitioners at large inexplicably ignore these problems.
However, a new perspective has emerged recently with the 
implementation~\cite{{GutierrezGuerrero:2010md},{Roberts:2010rn},{Roberts:2011wy}} of a 
confining, symmetry-preserving treatment of a vector-vector contact interaction as a 
representation of the gluon's two-point Schwinger function used in kernels of Dyson-Schwinger 
(DS) equations. By introducing a mechanism that ensures the absence of quark production thresholds, 
a feature of a confining theory, and embedding the interaction in a global-symmetry-preserving, 
rainbow-ladder truncation framework of the DS and BS 
equations~\cite{{Munczek:1994zz},{Bender:1996bb}}, the scheme has enhanced the capacity of a 
contact interaction to describe in a unified manner a diverse array of nonperturbative QCD 
phenomena -- for a recent review, see Ref.~\cite{Cloet:2013jya}

In this communication we examine the contact-interaction model within the perspective of 
a regularization scheme that allows to separate symmetry-offending parts in BS
amplitudes in a way independent of choices of momentum routing in divergent integrals, so that
the Ward-Green-Takahashi (WGT) identities reflecting global symmetries of the model are
preserved by the regularization. The 
scheme is inspired in a method introduced in Refs.~\cite{{Batt-thesis},{Battistel:1998tj}} 
to manipulate divergent Feynman integrals without specification of an explicit regulator. The 
method has been used with the NJL model in vacuum~\cite{Battistel:2008fd,Battistel:2013cja} 
and at finite temperature and baryon density~\cite{Farias:2006cs,Farias:2005cr,Farias:2007zz}. 
We formulate the problem generically with the aim of using it for heavy-light mesons, and
assess its reliability presenting numerical results for the $\pi$ and $K$ mesons.
 
\section{DS and BS equations and WGT identities}

We start with a brief review of the basic elements of the contact-interaction scheme 
introduced in Ref.~\cite{GutierrezGuerrero:2010md}. We consider the inhomogeneous BSE 
for a quark and antiquark state of total momentum $P$ (here and in the following we 
omit renormalization constants)  
\begin{eqnarray}
\hspace{-0.35cm}
\left[\Gamma_{\cal M}(k;P)\right]_{AB} &=& {\cal M}_{AB} 
+ \int_q \left[K(k,q;P)\right]_{AC,DB}
 \left[S(q_+) \Gamma_{\cal M}(q;P) S(q_-)\right]_{CD},
\label{eq:inhBSE}
\end{eqnarray}
where $\int_q\equiv\int d^4 q/(2\pi)^4$,  ${\cal M}$ gives the Dirac spinor structure of the state, $K(q,k;P)$ is the fully 
amputated quark-antiquark scattering kernel; $A,B,\cdots $ denote collectively color, flavor, 
and spinor indices; $q_\pm = q  \pm \eta_\pm P$, with $\eta_+ + \eta_- = 1$ and $q$ is the 
relative momentum. $S(k)$ is the dressed-quark propagator given by a DSE; for a given 
flavor~$f$ the general form of this DSE is
\begin{eqnarray}
S_f(k)^{-1} = i\gamma\cdot k + m_f 
+ \int_q \, g^2 
D_{\mu\nu}(k-q) 
 \, \frac{\lambda^a}{2}\gamma_\mu 
S_f(q) \frac{\lambda^a}{2}\Gamma^f_\nu(q,k),
\label{eq:DSEqp}
\end{eqnarray}
where $m_f$ is the current-quark mass. Here we will be interested in the flavor-nonsinglet
axial-vector $\Gamma^{lh}_{5\mu}(k;P)$ and pseudoscalar $\Gamma^{lh}_{\rm PS}(k;P)$
amplitudes for a quark of flavor $l$ and antiquark of flavor
$h$, for which ${\cal M}_{5\mu} = \gamma_5 \gamma_\mu$ and ${\cal M}_{PS} = \gamma_5$
respectively. Translation invariance requires that physical observables cannot depend on the 
choice of the momentum routing in quark propagators in Eq.~(\ref{eq:inhBSE}); that is, 
physical results must be independent of $\eta_\pm$. 

Associated with $\Gamma^{lh}_{5\mu}$ is the Ward-Green-Takahashi (WGT) identity
\begin{eqnarray}
P_\mu \Gamma_{5\mu}^{lh}(k;P) = S_l^{-1}(k_+) i \gamma_5 +  i \gamma_5 S_h^{-1}(k_-) 
 -  \, i\,(m_l + m_h) \,\Gamma^{lh}_{\rm PS}(k;P)\,,
\label{eq:avWGT}
\end{eqnarray}
where $\Gamma^{lh}_{PS}(k;P)$ is the pseudoscalar vertex. Pseudoscalar meson bound states 
are obtained from the solution of the homogeneous equation for $\Gamma^{lh}_{\rm PS}(k;P)$ 
\begin{eqnarray}
\left[\Gamma^{lh}_{\rm PS}(k;P)\right]_{AB} = \int_q\left[K(k,q;P)\right]_{AC,DB} 
\left[S_l(q_+)\Gamma^{lh}_{\rm PS}(q;P)S_h(q_-)\right]_{CD} ,
\label{eq:BSEps}
\end{eqnarray}
where here $A,B, \cdots$ denote color and spinor indices only. The general form of the of 
$\Gamma^{lh}_{\rm PS}(k;P)$ is
\begin{eqnarray}
\Gamma^{lh}_{\rm PS}(k,P) = \gamma_5 \bigl[ i E^{lh}_{\rm PS} 
+ \gamma\cdot P \, F^{lh}_{\rm PS} + \gamma\cdot k \, G^{lh}_{\rm PS} 
+ \, \sigma_{\mu\nu} k_\mu P_\nu \, H^{lh}_{\rm PS} \bigr],
\label{GammaPS}
\end{eqnarray}
where $E^{lh}_{\rm PS}, F^{lh}_{\rm PS}, \cdots$ are functions of $k$, $P$ and $k\cdot P$. 
The meson mass, $m_{\rm PS}$, is the eigenvalue $P^2 = - m^2_{\rm PS}$ that solves
Eq.~(\ref{eq:BSEps}). 

The contact-interaction scheme introduced in~Ref.~\cite{GutierrezGuerrero:2010md} 
amounts to make the following replacements in Eq.~(\ref{eq:DSEqp}) and the construction 
of scattering kernels of all BSEs
\begin{equation}
g^{2} D_{\mu\nu}(k-q) \rightarrow \frac{4\pi\alpha_{IR}}{m^2_G}\delta_{\mu\nu} 
\ \hspace{0.5cm}{\rm and}\hspace{0.5cm} 
\Gamma^a_\mu \rightarrow \frac{\lambda^a}{2}\gamma_\mu\,,
\label{eq:contact}
\end{equation}
where $m_G$ is a gluon mass-scale and $\alpha_{IR}$ is a fitting parameter. This means that 
the scattering kernel in all BSEs is 
\begin{equation}
\left[K(k,q;P)\right]_{AC,DB} = - \frac{4\pi\alpha_{IR}}{m^2_G}
\left(\frac{\lambda^a}{2}\gamma_\mu\right)_{AC} 
\left(\frac{\lambda^a}{2}\gamma_\mu\right)_{DB}  .
\label{eq:Kcontact}
\end{equation}
A feature of the momentum independence of the contact interaction is that the DS and BS equations
become nonrenormalizable.  This means that mass-scale parameters introduced with the regularization 
of divergent integrals cannot be removed from the calculations and need to be fixed
phenomenologically. Another feature, as mentioned previously, is 
that regularization of divergent integrals carries the danger of symmetry violation, in that 
WGT identities, like the one in Eq.~(\ref{eq:avWGT}), are not maintained, even when 
Poincar\'e-invariant regularization schemes are employed. 

Let us consider the DSE and the homogeneous pseudoscalar BSE, Eqs.~(\ref{eq:DSEqp})
and~(\ref{eq:BSEps}), with the contact interaction. In this case, $S_f(k)^{-1} = 
i\gamma\cdot k + M_f$, with
\begin{eqnarray}
M_f = m_f + \frac{64\pi\alpha_{IR}}{3m^2_G} \, 
\int_q \frac{M_f}{q^2+M^2_f}\,. 
\label{gapMf}
\end{eqnarray}
In addition, $G^{lh}_{\rm PS} = H^{lh}_{\rm PS} = 0$, 
so the Bethe-Salpeter amplitude Eq.~\eqref{GammaPS} takes the form 
\begin{eqnarray}
\label{contactBSA}
\Gamma^{lh}_{\rm PS}(P)=\gamma_5\left[iE^{lh}_{\rm PS}(P)+\frac{1}{2M_{lh}}\gamma\cdotp 
P\,F^{lh}_{\rm PS}(P)\right]~,
\end{eqnarray}
where $M_{lh}=M_lM_h/(M_l+M_h)$. Then, the  BSE can be written in the matrix form 
\begin{eqnarray}
\label{MHSE_mps}
\left[
\begin{array}{c}
E^{lh}_{\rm PS}(P)\\[0.2true cm]
F^{lh}_{\rm PS}(P)
\end{array}
\right]
&=& \frac{4\pi\alpha_{IR}}{3m^2_G}
\left[
\begin{array}{cc}
{\cal K}^{EE}_{lh}\;\; & {\cal K}^{EF}_{lh} \\[0.2true cm]
{\cal K}^{FE}_{lh}\;\; & {\cal K}^{FF}_{lh}
\end{array}\right]
\left[\begin{array}{c}
E^{lh}_{\rm PS}(P)\\[0.2true cm]
F^{lh}_{\rm PS}(P)
\end{array}
\right],\hspace{0.5cm}
\end{eqnarray}
where 
\begin{eqnarray}
{\cal K}^{EE}_{lh} &=& - \int_q\,
{\rm Tr}[\gamma_5 \, \gamma_\mu \, S_l(q_+) \, \gamma_5 \, S_h(q_-) \, \gamma_\mu]~,
\label{EE} \\[0.2true cm]
{\cal K}^{EF}_{lh} &=& \frac{i}{2M_{lh}} \int_q\,
{\rm Tr}[\gamma_5 \, \gamma_\mu \, S_l(q_+) \, \gamma_5 \, {\gamma\cdot P} \, 
S_h(q_-) \, \gamma_\mu]~, 
\label{EF} \\[0.2true cm] 
{\cal K}^{FE}_{lh} &=& \frac{2M^2_{lh}}{P^2}{\cal K}^{EF}_{lh} 
\label{FE} \\[0.2true cm]
{\cal K}^{FF}_{lh} &=& \frac{1}{P^2}\int_q
{\rm Tr}[\gamma_5 \,{\gamma\cdot P} \,\gamma_\mu  \,S_l(q_+)\gamma_5 
 \,{\gamma\cdot P} \, S_h(q_-) \,\gamma_\mu] .\hspace{1.9cm}
\label{FF}
\end{eqnarray}
Here, the trace is over Dirac indices. All integrals in Eqs.~(\ref{gapMf}) and 
(\ref{EE})-(\ref{FF}) are ultraviolet divergent; the divergences are quadratic 
and logarithm. The vast majority of applications within NJL models ignore the 
pseudo vector component $F^{lh}_{\rm PS}(P)$; in doing so, this leads to the 
random-phase-approximation (RPA) for the BS equation. 

\section{Symmetry-preserving subtraction scheme }
The issue of symmetry violation can be exposed examining the momenta running in the quark 
propagators; they are $q_+ = q + k_1$ and $q_- = q + k_2$, with $k_1 = \eta_+ P$ and $k_2 
= - \eta_- P$. The momenta $k_1$ and $k_2$ are arbitrary, as while satisfying $\eta_+ 
+ \eta_- = 1$, the $\eta_\pm$ are otherwise arbitrary. However, to maintain translation 
invariance, the integrals must depend only on the relative momentum $k_1 - k_2$ or, 
equivalently, the integrals must not depend on $\eta_\pm$, they can depend only in the 
combination $\eta_+ + \eta_- = 1$. The dependence on $k_1 - k_2$ only is also crucial for 
maintaining the WGT identities, like the one in Eq.~(\ref{eq:avWGT}). 

The traditional way of handling within NJL-type of models integrals like those in 
Eqs.~(\ref{EE})-(\ref{FF}) is as follows: after evaluating the traces, a choice
for $\eta_+$ and $\eta_-$ is made and Feynman parameters are used to combine in a 
single term the product $(q^2_+ + M^2_l)(q^2_- + M^2_h)$ in the denominator; then, after a 
{\em momentum shift} to eliminate the angle $q \cdot P$ in the denominator, the integral 
over $q$ is performed. In making the momentum shift, changes in the integration limits are 
ignored. Invariably, results depend upon the choices made for $\eta_\pm$; 
in particular, the value of the leptonic decay constants, which play a key role in 
the normalisation of the pion BSE, depend upon the choices made for $\eta_\pm$. Some 
regularization-independent results, like the Goldberg-Treiman relation at the quark level 
and the Gell-Mann--Oakes-Renner relation, can be obtained after using the gap 
equation to eliminate the quadratic divergences. 

The subtraction scheme of Refs.~\cite{Batt-thesis,Battistel:1998tj,Battistel:2008fd,
Battistel:2013cja,Farias:2006cs,Farias:2005cr,Farias:2007zz} is based on the repeated 
use of the identity
\begin{eqnarray}
\frac{1}{q^2_\pm + M^2} &=& \left(\frac{1}{q^2_\pm + M^2} - \frac{1}{q^2 + M^2}\right) 
+ \frac{1}{q^2 + M^2} \nonumber \\[0.25true cm]
&=&  \frac{1}{q^2 + M^2} - \frac{q^2_\pm - q^2 }
{\left(q^2 + M^2\right)}\frac{1}{\left(q^2_\pm + M^2\right)}.
\label{subtr-1}
\end{eqnarray}  
Assuming a Poincar\'e invariant regularization for the 
integrals in Eqs.~(\ref{EE})-(\ref{FF}), subtractions are performed for each of the 
propagators $S_l(q_+)$ and $S(q_-)$, the number of which is dictated by the requirement 
that a finite integral is obtained - note that while the original denominator goes as 
$1/q^2$ for $q \rightarrow \infty$, the last term in Eq.~(\ref{subtr-1}) 
goes as $1/q^3$ for $q \rightarrow \infty$. We illustrate the procedure in detail for 
the ${\cal K}^{EE}_{lh}(P)$ kernel. Evaluation of the trace in Eq.~(\ref{EE}) leads to 
\begin{eqnarray}
\hspace{-0.25cm}{\cal K}^{EE}_{lh} &=&  16 \int^\Lambda_q
\frac{q_+ \cdot q_- +M_l M_h}{(q^2_+ + M^2_l)(q^2_- + M^2_h)},
\label{EE-tr}
\end{eqnarray}
where $\Lambda$ indicates the mass scale associated with the regularization. Next, subtracting
thrice each of the denominators, ${\cal K}^{EE}_{lh}$ can be written as a sum of three kinds of 
terms: quadratic and logarithmically divergent integrals that are independent of $\eta_\pm$, 
symmetry violating terms that are proportional to  $\eta_+$ and $\eta_-$, and a finite 
integral also independent of $\eta_\pm$. Explicitly, they are given by the following expressions:
\begin{eqnarray}
{\cal K}^{EE}_{\rm PS} &=&  8 \Bigl\{
\left[\eta^2_+ \, A_{\mu\nu}(M^2_l) + \eta^2_- \, A_{\mu\nu}(M^2_h) \right] 
P_\mu P_\nu + \, I_{\rm quad}(M^2_l)  + I_{\rm quad}(M^2_h) 
\nonumber \\[0.25true cm]
&& - \left[ P^2 + \left(M_h - M_l\right)^2 \right] \,
\left[ I_{\rm log}(M^2_l) - Z_0(M^2_l,M^2_h,P^2;M^2_l) \right]\Bigr\},
\label{EE-fin}
\end{eqnarray}
where $A_{\mu\nu}(M^2)$, $I_{\rm quad}(M^2)$ and $I_{\rm log}(M^2)$ are the divergent integrals (when $\Lambda \rightarrow \infty$)
\begin{equation}
A_{\mu\nu}(M^2) = \int^\Lambda_q  
\frac{ 4q_\mu q_\nu - (q^2+M^2) \delta_{\mu\nu} }{ (q^2+M^2)^3 },
\label{A_munu}
\end{equation}
\begin{eqnarray}
I_{\rm quad}(M^2) = \int^\Lambda_q \,  \frac{1}{q^2+M^2},
\hspace{1.5cm}
I_{\rm log}(M^2) = \int^\Lambda_q \, \frac{1}{(q^2+M^2)^2}~,
\label{Iquad-Ilog}
\end{eqnarray} 
and $Z_0(M^2_l,M^2_h,P^2;M^2_l)$ is a finite integral given by
\begin{eqnarray}
Z_0(M^2_l,M^2_h,P^2;M^2_l) = \frac{1}{(4\pi)^2}\int^1_0\,dz\, 
\ln\left[\frac{P^2z(1-z)-(M^2_l-M^2_h)z+M^2_l}{M^2_l}\right].
\end{eqnarray}
It is important to note that to arrive at these results, no momentum shifts were made 
in divergent integrals. The other amplitudes,
${\cal K}^{EF}_{lh}$, ${\cal K}^{FE}_{lh}$, and ${\cal K}^{FF}_{lh}$ are given by similar 
integrals~\cite{SABK}.

Note that Eq.~(\ref{EE-fin}) makes it clear that whatever choice made for $\eta_\pm$,  
unavoidably translation symmetry is broken, {\em unless} the regularization scheme 
leads to $A_{\mu\nu}(M^2) = 0$. Let us next examine how choices of $\eta_{\pm}$ lead to
violation of the WGT identity in Eq.~(\ref{eq:avWGT}). One needs to deal with integrals of 
the form
\begin{eqnarray}
\int^\Lambda_q \frac{q_\mu }{q^2_\pm + M^2} \! &=& \!
\mp \, \eta_\pm P_\mu \, I_{\rm quad}(M^2) \pm \, \frac{1}{3} \eta^3_\pm 
\left[P^2 A_{\mu \rho}(M^2) P_\rho 
- P_\mu A_{\rho \sigma}(M^2) P_\rho P_\sigma\right] \nonumber \\[0.25true cm]
&& \mp \, \eta_\pm B_{\mu \rho}(M^2) P_\rho \mp \frac{1}{3} \eta^3_{\pm}
 C_{\mu \rho \sigma \lambda}(M^2) P_\sigma P_\rho P_\lambda , 
\label{WGT-ABC}
\end{eqnarray}
where $B_{\mu\nu}(M^2)$ and $C_{\mu\nu\rho\sigma}(M^2)$ are the new structures
\begin{eqnarray}
B_{\mu\nu}(M^2) = \int^\Lambda_q  
\frac{2q_\mu q_\nu - (q^2+M^2)\delta_{\mu\nu}}{(q^2+M^2)^2},
\hspace{1.0cm}
C_{\mu\nu\rho\sigma}(M^2) = \int^\Lambda_q  
\frac{ c_{\mu \nu \rho \sigma}(q,M^2)}{(q^2+M^2)^4}  ,
\label{B&C}
\end{eqnarray}
with
\begin{eqnarray}
c_{\mu\nu\rho\sigma}(q^2,M^2) = 24 q_\mu q_\nu q_\rho q_\sigma 
- \, 4 (q^2+M^2)  ( \delta_{\mu\nu} q_\rho q_\sigma 
+ {\rm perm.} \; \nu\sigma\rho) .
\label{cmunusr}
\end{eqnarray}
Using these integrals in the WGT identity, Eq.~(\ref{eq:avWGT}), one obtains
\begin{eqnarray}
0 = (M_l - m_l) + (M_h - m_h) 
 - \, \frac{64\pi\alpha_{IR}}{3m^2_G}\bigg\{ I_{\rm quad}(M^2_l) 
+ I_{\rm quad}(M^2_h) + \left[\eta^2_+ \, A_{\mu\nu}(M^2_l) + \eta^2_- \, A_{\mu\nu}(M^2_h)\right] \bigg\} P_\mu P_\nu ,
\label{WGT-gap}
\end{eqnarray}
and
\begin{eqnarray}
0 = \int^\Lambda_q 
\left(\frac{q_+\cdot P}{q^2_+ + M^2}  - \frac{q_-\cdot P}{q^2_- + M^2}\right) 
\sim  {\rm terms}\;\;{\rm proportional}\;\; {\rm to} \;\; \eta_\pm  \left(A_{\mu \nu}, B_{\mu \nu},
C_{\mu\nu\rho\sigma}\right).
\label{int-ABC}
\end{eqnarray}
We see that for arbitrary momentum routing in the loop
integrals, i.e. for arbitrary values of $\eta_{\pm}$ satisfying $\eta_+ + \eta_-=1$, the 
subtraction scheme allows to identify in a systematic way symmetry offending terms; they 
are the integrals $A_{\mu\nu}$, $B_{\mu\nu}$ and $C_{\mu\nu\rho\sigma}$ in the equations above. 
A consistent regularization scheme should make the integrals vanish automatically. Otherwise, 
the vanishing of the integrals must be imposed; in doing so, the regularization scheme becomes
part of the model. Dimensional regularization and Pauli-Villars regularization are examples
of schemes that lead to $A_{\mu\nu}=0$, $B_{\mu\nu} = 0$, and $C_{\mu\nu\rho\sigma} = 0$. 
Removing the symmetry offending terms, the kernels ${\cal K}^{EE}_{lh}$, 
${\cal K}^{EF}_{lh}$, ${\cal K}^{FE}_{lh}$, and ${\cal K}^{FF}_{lh}$ become
free from ambiguities and symmetry-preserving.

Let us make contact with the scheme of Ref.~\cite{GutierrezGuerrero:2010md}. For 
$M_l = M_h = M$ and $\eta_+ = 1$ and $\eta_- = 0$, which is the choice made in 
Ref.~\cite{GutierrezGuerrero:2010md}, Eq.~(\ref{int-ABC}) becomes
\begin{eqnarray}
\int^\Lambda_q 
\left(\frac{q_+\cdot P}{q^2_+ + M^2}  - \frac{q\cdot P}{q^2 + M^2}\right) = 0.
\label{WGT-MM}
\end{eqnarray}
This is precisely Eq.~(15) of Ref.~\cite{GutierrezGuerrero:2010md}. Using Eq.~(\ref{WGT-ABC}) 
for this integral, one obtains
\begin{eqnarray}
\hspace{-0.0cm}0 = \int^\Lambda_q 
\left(\frac{q_+\cdot P}{q^2_+ + M^2}  - \frac{q\cdot P}{q^2 + M^2}\right) 
=  P_\mu \bigl[ B_{\mu \nu}(M^2) + \frac{1}{3} C_{\mu\nu\rho\sigma}(M^2) P_\rho P_\sigma 
+   \frac{1}{3}P_\mu A_{\rho \nu }(M^2) P_\rho 
- \frac{4}{3}P^2 A_{\mu\nu}(M^2) \bigr] P_\nu ,
\end{eqnarray}
that is
\begin{eqnarray}
B_{\mu \nu}(M^2) + \frac{1}{3} C_{\mu\nu\rho\sigma}(M^2) P_\rho P_\sigma 
 + \, \frac{1}{3} P_\mu A_{\rho \nu }(M^2) P_\rho 
-\frac{4}{3}P^2 A_{\mu\nu}(M^2) = 0.
\end{eqnarray}
In the quiral limit, $P=0$, this leads to
\begin{equation}
B_{\mu\nu}(M^2) = 0 = \delta_{\mu\nu} \int_\Lambda \frac{d^4 q}{(2\pi)^4}
\frac{\frac 1 2 q^2 + M^2}{(q^2 + M^2)^2},
\end{equation}
which is Eq.~(17) of Ref.~\cite{GutierrezGuerrero:2010md}. This result makes
it clear that our scheme, besides being in agreement with Ref.~\cite{GutierrezGuerrero:2010md}
for the particular choice of $\eta_\pm$, it is also more general, as it is valid 
for {\em arbitary} values $\eta_\pm$.

\section{Numerical results}

In order assess the reliability of the subtraction scheme, we apply it to the calculation 
of the masses and electroweak decay constants of the $\pi$ and $K$ mesons. 
The mass of the meson is obtained from Eq.~\eqref{MHSE_mps}; it is an
eigenvalue problem which has a solution at $P^2 = -m^2_{\rm PS}$. We use the canonical 
normalization condition for the BS amplitudes, namely
\begin{eqnarray}
1 = \frac{\partial}{\partial P^2}\Pi^{lh}_{\rm PS}(Q,P)\bigg|_{Q=P} ,
\end{eqnarray}
with 
\begin{eqnarray}
\label{Pi_nor}
\Pi^{lh}_{\rm PS}(Q,P) = 6\int\frac{d^4q}{(2\pi)^4}{\rm {Tr}}[\bar\Gamma^{hl}_{\rm PS}(-Q)S_l(q+k_1)
\Gamma^{lh}_{\rm PS}(Q)S(q+k_2)]~, 
\end{eqnarray}
where $\bar\Gamma^{hl}_{\rm PS}(-Q)=\left[C^{-1}\Gamma^{lh}_{\rm PS}(-Q)C\right]^T
= \Gamma^{lh}_{\rm PS}(-Q)$. The electroweak decay constant of the meson, $f_{\rm PS}$, 
can be expressed in terms of normalized amplitudes $E$ and $F$ in the form 
\begin{eqnarray}
f_{\rm PS} = \frac{6}{M_{lh}}\left[E_{\rm PS}(P){\cal K}^{FE}_{\rm PS}
+F_{\rm PS}(P){\cal K}^{FF}_{\rm PS}\right]\bigg|_{P^2=-m^2_{\rm PS}}.
\end{eqnarray}

We compare results with Ref.~\cite{Chen:2012txa}, where the scheme of 
Ref.~\cite{GutierrezGuerrero:2010md} was used for the $K$ meson. We use proper-time 
regularization for $I^{\Lambda}_{quad}(M_f)$. In the BS amplitudes where 
$I^{\Lambda}_{quad}(M_f)$ appears, we use the gap equation \eqref{gapMf}, and 
for $I^{\Lambda}_{log}(M_f)$ use the identity 
\begin{equation}
I^{\Lambda}_{log}(M_f) = - \frac{d I^{\Lambda}_{quad}(M_f)}{d M^2_f}.
\end{equation}
On the other hand, for the subtraction mass scale we have used the light mass $M^2_l$ 
and finite integrals like $Z_0(M^2_h,M^2_l,P^2;M^2_h)$ are performed  without imposing
a regulator.

We use the same parameters employed in Ref.~\cite{Chen:2012txa}: $\alpha_{IR}=0.93\pi$,  
$m_G=0.8$~GeV, $m_u=m_d=0.007$~GeV and $m_s=0.160$~GeV, $\tau^2_{IR}=(0.24\, {\rm GeV})^2$ 
and $\tau^2_{UV}=(0.905\, {\rm GeV})^2$. The last two parameters are the infrared and 
ultraviolet cutoff parameters of the proper-time regularization. The solution of the gap 
equation leads for the constituent quark masses $M_u = M_d = 0.367$~GeV, and $M_s=0.525$~GeV.   
Table~I presents our results and those of Ref.~\cite{Chen:2012txa} (indicated with a star). 

\vspace{0.5cm}
\begin{table}[ht]
\caption{Masses and electroweak decay constants$\pi$ and $K$ mesons. 
The four first columns show the results obtained using the subtraction scheme and 
the four last indicated with a star, are the results of Ref.~\cite{Chen:2012txa}.}
\begin{tabular}{cc|cccc|cccc}
\hline
& \tablehead{1}{c}{c}{Meson}
 & \tablehead{1}{c}{c}{$E$}
& \tablehead{1}{c}{c}{$F$}
  & \tablehead{1}{c}{c}{ $m_{\rm PS}$ }
  & \tablehead{1}{c}{c}{$f_{\rm PS}$}   
    & \tablehead{1}{c}{c}{$E^*$}
& \tablehead{1}{c}{c}{$F^*$}
  & \tablehead{1}{c}{c}{ $m^*_{\rm PS}$ }
  & \tablehead{1}{c}{c}{$f^*_{\rm PS}$} \\
\hline
& $\pi$ & 3.759 & 0.498 & 0.139 & 0.106 & 3.596 & 0.474 & 0.139 &0.101\\
& $K$   & 3.984 & 0.632 & 0.494 & 0.115 & 3.864 & 0.591 & 0.493 &0.107\\
\hline
\end{tabular}
\label{tab:a}
\end{table}

\vspace{0.5cm}

We see from Table~I that the results obtained with the subtraction scheme compares well
with those of Ref.~\cite{Chen:2012txa}: the masses are almost identical in both approaches,
and the decay constants from the subtraction are a little larger, but by less than 5\%.
We recall that the differences in the approaches are that in the subtraction scheme the
results are independent of the choice of $\eta_\pm$, while those of Ref.~\cite{Chen:2012txa}
are for $\eta_+ = 1$ and $\eta_- =0$. Another difference is that the finite integral 
$Z_0(M^2_h,M^2_l,P^2;M^2_h)$ is integrated without imposing ultraviolet or
infrared cutoffs. There is no difficulty in using an infrared cutoff in the finite integrals
to avoid unphysical quark-antiquark thresholds in amplitudes.

\section{Conclusions and Perspectives}

We examined the contact-interaction model introduced in Ref.~\cite{GutierrezGuerrero:2010md}
within the perspective of a regularization scheme that allows to separate symmetry-offending 
parts in Bethe-Salpeter amplitudes in a way independent of choices of momentum routing in 
divergent integrals. In doing so, the Ward-Green-Takahashi (WGT) identities reflecting 
global symmetries of the model are preserved by the regularization. Symmetry-offending
parts of the amplitudes, the integrals $A_{\mu\nu}$, $B_{\mu\nu}$ and $C_{\mu\nu}$,
can be neatly separated. In general, a cutoff regularization scheme leads to nonzero values 
for the symmetry-offending integrals, while Pauli-Villars or dimensional regularization lead 
to the vanishing of the symmetry-offending integrals. In a nonrenormalizable model, like the
contact-interaction model discussed here, the vanishing of $A_{\mu\nu}$, $B_{\mu\nu}$ and 
$C_{\mu\nu}$ must be imposed in an {\em ad hoc} manner.

Our aim is to apply the subtraction scheme to heavy-light mesons, where the heavy quark
is much heavier than the light quark, like in the $D$ and $B$ mesons. When the masses are 
not much different, one can show that in the RPA approximation to the BS equation, 
the heavy-light pseudoscalar meson mass $m^2_{\rm PS}$ and corresponding 
electroweak decay constant $f_{\rm PS}$ can be expressed in terms of the light-quark 
condensate $\langle \bar q q\rangle$ and the pion decay constant $f_\pi$ as 
\begin{eqnarray}
m^2_{\rm PS} & \simeq & (M_h - M_l)^2 - m_h \frac{M_l}{M_h} 
\frac{\langle \bar q q\rangle}{f^2_\pi} \label{mPS} \\[0.2true cm]
f_{\rm PS} & \simeq & \frac{1}{2}\left(1 + \frac{M_h}{M_l}\right) f_\pi 
\label{fPS}
\end{eqnarray}  
These expressions are valid for $M_h \leq \Lambda$, where $\Lambda$ is the 
regularization mass scale in the divergent integrals $I_{\rm quad}$ and $I_{\rm log}$. 
Eq.~(\ref{fPS}) shows very clearly that $f_\pi < f_K < f_D$, which is the correct ordering
of their experimental values. Using the traditional cutoff regularization scheme, in which 
no attention is given to symmetry-offending terms in amplitudes, one obtains~\cite{blaschke}
$f_\pi \simeq f_K$ and $f_D < f_\pi$. Of course, Eqs.~(\ref{mPS}) and (\ref{fPS}) are 
not strictly applicable to this case, but they show the trend one can expect when using
the subtraction scheme. Because of the large diferences between the masses of the quarks 
running in the loops, the application of the subtraction method in this case is 
nontrivial and needs adaptations -- results will be presented elsewhere~\cite{SABK}.

\begin{theacknowledgments}
The work of M.A.B. and F.E.S. were financed by doctoral scholarships by Coordena\c{c}\~ao
de Aperfei\c{c}oamento de Pessoal de N\'{\i}vel Superior - CAPES and Conselho Nacional de Desenvolvimento Cient\'{\i}fico e Tecnol\'ogico - CNPq, respecytively. The work of G.K. was 
supported in part by Conselho Nacional de Desenvolvimento  Cient\'{\i}fico e Tecnol\'ogico 
- CNPq, Grant No. 305894/2009-9, and Funda\c{c}\~ao de Amparo \`a Pesquisa do Estado de 
S\~ao Paulo - FAPESP, Grant No. 2013/01907-0. 
\end{theacknowledgments}

\bibliographystyle{aipproc}   
\bibliographystyle{aipprocl} 

\bibliography{sample}

\IfFileExists{\jobname.bbl}{}
 {\typeout{}
  \typeout{******************************************}
  \typeout{** Please run "bibtex \jobname" to optain}
  \typeout{** the bibliography and then re-run LaTeX}
  \typeout{** twice to fix the references!}
  \typeout{******************************************}
  \typeout{}
 }

\end{document}